# Reconstruction of energy and angle distribution function of surface-emitted negative ions in hydrogen plasmas using mass spectrometry

D. Kogut[1], K. Achkasov[1], J. P. J. Dubois[1], R. Moussaoui[1], J. B. Faure[1], J. M. Layet[1], A. Simonin[2], G. Cartry[1]

[1]*Aix-Marseille Université, CNRS, PIIM, UMR 7345, 13397 Marseille, France*
[2]*CEA, IRFM, F-13108 Saint-Paul-lez-Durance, France*

## Abstract

A new method involving mass spectrometry and modelling is described in this work, which may highlight the production mechanisms of negative ions on surface in low pressure plasmas. Positive hydrogen ions from plasma impact a sample which is biased negatively with respect to the plasma potential. Negative ions (NI) are produced on the surface through the ionization of sputtered and backscattered particles and detected according to their energy and mass by a mass spectrometer placed in front of the sample. The shape of the measured negative-ion energy distribution function (NIEDF) strongly differs from the NIEDF of the ions emitted by the sample because of the limited acceptance angle of the mass spectrometer. The reconstruction method proposed here allows to compute the distribution function in energy and angle (NIEADF) of the negative-ions emitted by the sample based on the NIEDF measurements at different tilt angles of the sample. The reconstruction algorithm does not depend on the NI surface production mechanism, so it can be applied to any type of surface and/or NI. The NIEADFs for HOPG (Highly Oriented Pyrolitic Graphite) and Gadolinium (low work-function metal) are presented and compared with the SRIM modelling. HOPG and Gd show comparable integrated NI yields, however the key differences in mechanisms of NI production can be identified. While for Gd the major process is backscattering of ions with the peak of NIEDF at 36 eV, in case of HOPG the sputtering contribution due to adsorbed H on the surface is also important and the NIEDF peak is found at 5 eV.

Corresponding author: gilles.cartry@univ-amu.fr

# 1. Introduction

Negative-ions (NI) can be created in low-pressure plasmas either by dissociative attachment of electrons on molecules in the volume [1,2] or by conversion of positive ions or hyperthermal neutrals on the surface immersed into plasma [3,4]. Volume production of NI is widely applied in microelectronics industry [5] and space propulsion engines [6,7], while the surface production mechanism is essential in NI sources for magnetically confined fusion reactors [8,9] and particle accelerators [10,11]. NI can be also formed on the target surface as a byproduct during the reactive magnetron sputtering [12].

Deuterium NI generation is of a primary interest for tokamaks (magnetic confinement fusion), where a plasma-based $D^-$ source is used to produce an energetic ion beam which is neutralized through interaction with a gas stripper and is injected into the high-temperature plasma core of the reactor to provide heating and current drive. In case of the largest fusion device ITER, which is under construction now, a 40 A current of $D^-$ with energy of 1 MeV has to be extracted [13,14,15]. At these energies neutralization efficiency of $D^+$ tends to zero, while an extra electron can be easily detached from $D^-$ through collisions with gas [13] or by laser photo-detachment [16]. In order to meet requirements in terms of the NI current a cesium (Cs) evaporation in the source chamber is used, which increases strongly the NI surface conversion rate [17]. This technique complicates long-term operation of the reactor due to a number of drawbacks, such as high Cs consumption, possible Cs contamination and breakdowns in the accelerator stage; hence, development of an alternative NI source is a subject of intensive studies [18,19,20,21]. In this work a new experimental method to study NI surface production in cesium-free low-pressure hydrogen plasmas is proposed.

In our experimental device [20] a sample is introduced in the plasma chamber and negatively biased with respect to the plasma potential to attract the positive ions. Surface-produced NI are accelerated by the sheath in front of the sample and directed towards a mass spectrometer (MS), where they are detected according to their energy and mass. It has been shown that HOPG (Highly Oriented Pyrolitic Graphite) is a good negative-ion surface-production enhancer material when exposed to a low pressure hydrogen plasma [22,23]. Furthermore, this material can be easily cleaved, allowing for a repeated use of new fresh samples with identical properties to the previous ones; hence, HOPG is used as a reference material in the present studies.

In order to gain an insight on the mechanisms of the NI surface production, it is important to analyze and characterize the shape of the measured Negative-Ion Energy Distribution Function (NIEDF). It strongly differs from the NIEDF of the ions emitted by the sample because of the limited acceptance angle of the mass spectrometer, so a model has been developed previously to interpret the experimental results [20]. In addition, a method has been proposed [24] to obtain the distribution functions in energy and angle (NIEADFs) of the negative-ions emitted by the sample based on an *a priori* assumption of NIEADF given by the SRIM code [25] and its *a posteriori* validation by comparison of the modelled and experimental NIEDFs at different tilts of the sample. SRIM output includes both sputtered and backscattered particles, however it does not take into account the probability to form NI on the surface: $P_{iz}$. It has been suggested earlier that this probability is constant ($P_{iz} = const$) for HOPG, i.e. independent of the neutral particle energy and angle of emission, as a remarkable agreement of experimental NIEDFs with the SRIM-based modelling has been demonstrated [24]. However, if we consider other materials, the ionization probability $P_{iz}$ may not be constant. For instance, $P_{iz}$ on the surface of low work function metals may depend on the perpendicular velocity of the outgoing particle [26]. Furthermore, in order to obtain a correct NIEADF by SRIM the proper input parameters are needed, such as a surface binding energy, a surface concentration of hydrogen; these are known well for carbon owing to the decades of fusion research, however they may not be defined for other materials. Therefore, in general case, it is necessary to determine the distribution in energy and angle of NI emitted from the surface purely from the experimental data without any *a priori* assumptions about NIEADF, $P_{iz}$ or the input parameters of SRIM. Here we propose a new method which allows to reconstruct the full NIEADF at the sample surface based on the NIEDF measurements at different tilt angles of the sample. It does not depend on the NI surface production mechanism, so it can be applied to any type of surface and/or NI. We present the results for HOPG to check the validity of the method through comparison with the SRIM modelling; we also show NIEADF for gadolinium as an example of low work-function metal (2.9 eV [27]), which can be further implemented as an alternative to cesium.

## 2. Experimental set-up

The reactor, diagnostic instruments and plasma conditions used are described in detail elsewhere [20,24]. Measurements are performed in a spherical vacuum chamber (radius 100 mm), see Figure 1. The sample is introduced in the center with a molybdenum substrate holder that can be negatively biased. The plasma is created either with RF power (13.56 MHz) applied to a Boswell antenna on top of the chamber or with an ECR source (Electron Cyclotron Resonance, 2.45 GHz) from Boreal Plasma, which is installed at 5 cm away from the sample. The sample surface exposed to the plasma is a disc of 8 mm in diameter facing the mass spectrometer entrance located at 37 mm away. The latter consists of a 40 mm in diameter grounded cylinder with a 5 mm in diameter opening centered on its base, where the extractor electrode with a sampling orifice of 100 μm is installed, as shown in the inset in Figure 1. The mass spectrometer axis passes through the center of the sample and the latter can be rotated in the direction perpendicular to this axis (see Figure 1). Langmuir probe can be inserted in the center of the chamber to measure the plasma parameters.

The operation conditions in case of RF plasma are the following: 2 Pa $H_2$, 20 W of injected power. Such low level of power has been chosen to avoid which affect the NIEDF measurements; the discharge is operated in the capacitive coupling regime. There is a grounded metal plate 50 mm above the MS and the sample to reduce the perturbations of plasma potential (not shown in Figure 1, see details in [20]). The plasma density $n_e$, as measured by Langmuir probe, is in the range of $10^{13}$-$10^{14}$ m$^{-3}$ and the electron temperature is $T_e = 3.5$ eV. Due to the low electron density, there is a large uncertainty in these measurements; in addition, the average ion flux to the sample is small (of the order of $10^{17}$ m$^{-2}$s$^{-1}$), hence this discharge is less convenient for NI studies.

In case of ECR plasma 1 Pa H$_2$ and 60 W power is used; $n_e = 2.5 \cdot 10^{15}$ m$^{-3}$, $T_e = 1.0$ eV and the plasma potential is $V_p = 7$ V. The ECR plasma produces much higher ion flux to the sample compared to RF: ~$7 \cdot 10^{18}$ m$^{-2}$s$^{-1}$, which yields in higher NI intensities. Moreover, NIEDFs are not perturbed at the ECR frequency, which makes this regime particularly interesting for NI studies [24].

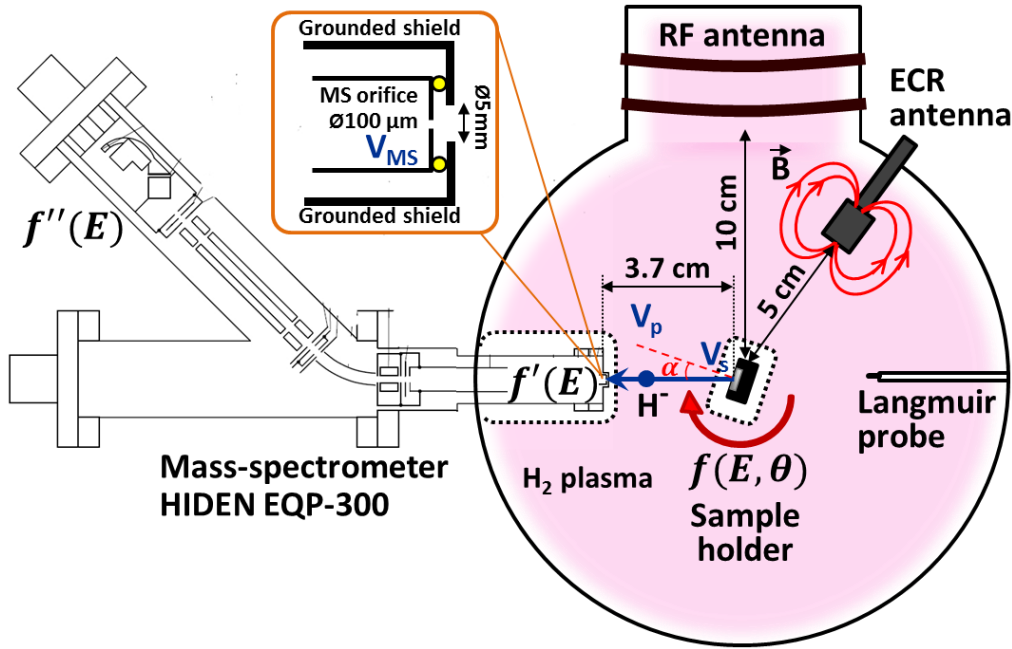

Figure 1. Schematic of the experimental set-up. White areas around the mass-spectrometer entrance and the sample holder represent sheaths. $V_p$ is the plasma potential, $V_s$ is the sample bias and $V_{MS}$ is the potential applied to the extractor electrode with a sampling orifice of the mass-spectrometer. The extractor itself is shielded with a grounded cylinder, as shown in the inset. Negative ions are emitted from the sample surface (shown in gray) with the distribution function $f(E, \theta)$ and a part of them is collected by the mass-spectrometer giving the distribution $f_{\exp}'(E)$ at the entrance and $f_{\exp}''(E)$ at the detector. Angle $\alpha$ stands for the tilt of the sample with respect to the MS axis.

Sample is biased at $V_s = -130$ V, negatively with respect to the plasma potential $V_p$. Negative ions are formed on the sample surface upon the positive ion bombardment and accelerated by the sheath towards the plasma. Under the low pressure conditions considered here, most negative ions cross the plasma without any collision (a mean free path for the electron detachment is in the range of 40–80 mm) [24]. Finally, a part of negative ion flux is collected by the mass spectrometer contributing to the Negative-Ion Energy Distribution Function (NIEDF) $f_{\exp}'(E)$; then the NIEDF measured by the MS detector is $f_{\exp}''(E)$. These functions are labeled with prime marks in contrast to the distribution $f_{\exp}(E)$ of all NI leaving the sample surface, which is different for the reasons given below. The MS extractor electrode with a sampling orifice can be biased at $V_{MS}$ (see Figure 1); during the NI measurement $V_{MS}$ is kept at 0 V to prevent the distortion of the planar sheath in front of the grounded MS entrance shield [20].

The microwave frequency of the ECR plasma is 2.45 GHz. It corresponds to a resonance with the cyclotron frequency of electrons for a magnetic field of 845 G reached in the vicinity of the antenna.

However, at the point where the sample holder is located, the magnetic field is less than 50 G; this is relatively small and should not affect the NI trajectories. Indeed, the Larmor radius for the NI accelerated by the sheath in front of the sample is more than 33 cm, which is much larger than the distance between the MS entrance and the sample. If the antenna is moved farther than 5 cm from the sample, the measured intensity gets smaller, but the shape of the NIEDF is not changed. As far as the plasma homogeneity in front of the sample is concerned, a model has shown that $n_e$ profile is homogeneous in the axial direction from a similar ECR antenna (no noticeable gradients on the scale of 1-4 cm) for 1 Pa Ar plasma at 50 W of absorbed power [28].

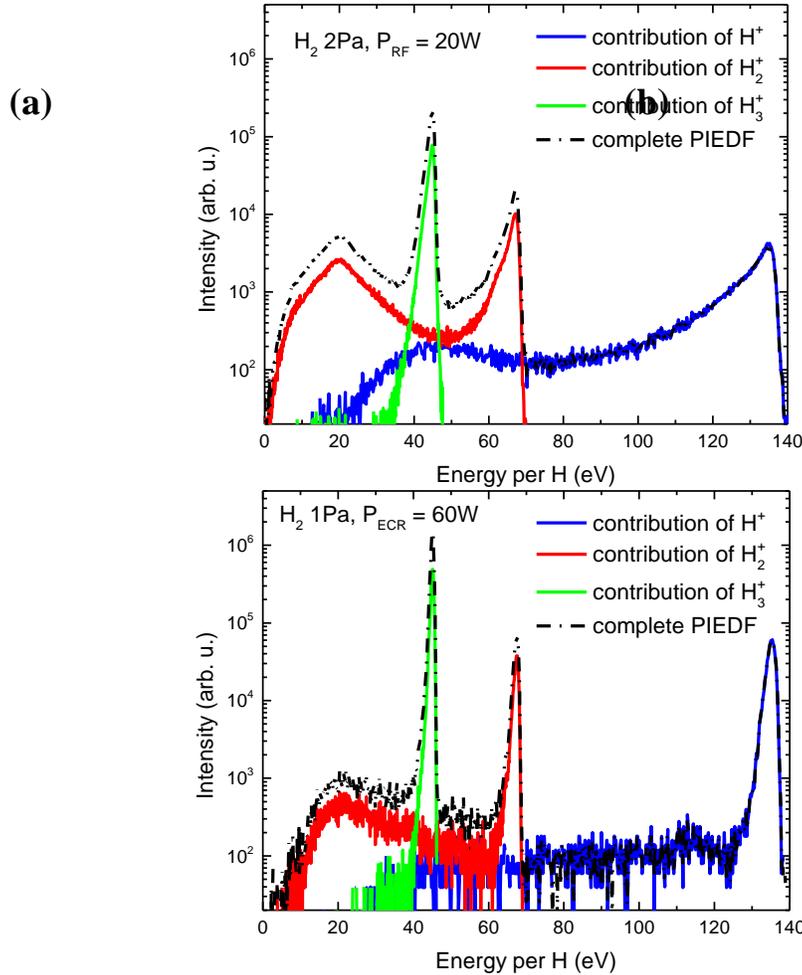

Figure 2. Positive-Ion Energy Distribution Function measured at $V_{MS}$ = -130 V and $V_s$ = 0 V for the RF H$_2$ plasma at 2 Pa, 20 W (a) and ECR H$_2$ plasma at 1 Pa, 60 W (b).

The positive ion flux composition is also determined by mass spectrometry. Positive ions cross a high voltage sheath before impinging the sample. Therefore, in order to assess correctly the Positive-Ion Energy Distribution Function (PIEDF) at the sample surface, the mass-spectrometer sampling orifice is polarized at $V_{MS}$ = -130 V and the sample holder is grounded. The measured intensities are corrected for the number of protons and the energy per proton for each type of ion; the resulting total PIEDFs for RF and ECR plasmas are shown in Figure 2. It is assumed that H$_2^+$ and $H_3^+$ are immediately dissociated at the sample surface and the kinetic energy is equally distributed between the impinging protons. It should be noted though that due to the variation of the transmission probability of the mass spectrometer with ion mass and energy [4,20,29] these are approximate PIEDFs: the measured intensities can vary by few percent depending on the mass spectrometer tuning. One may note substantial differences between two regimes in Figure 2: the peaks are more pronounced and demonstrate higher intensities in case of the ECR plasma. This is due to higher $n_e$ and thinner sheaths: positive ions experience fewer collisions on their way through the sheath, so their energy distribution is less perturbed. On the contrary, in case of RF one can observe a substantial increase of low-energy part of the distribution of $H_2^+$ and $H^+$, which does not occur for $H_3^+$. The low-energy ions $H_2^+$ and $H^+$ are the products of charge-exchange reactions in the sheath, while the reaction $H_2^+ + H_2 \rightarrow H_3^+ + H$ is not probable as the sheath dimensions are smaller than the corresponding mean free path [30]. $H^+$ can also loose energy through the ro-vibrational excitation of the $H_2$ molecules. Charge-exchange reactions produce fast H atoms, which may contribute to the surface formation of $H^-$; this process is not taken into account in the present model. However, let us note that the PIEDFs taken into account in the present paper are more realistic than the mono-energetic ones taken into account in the previous papers [20,24] and improve the validation of the modelling method. As far as integrals of PIEDF are concerned, $H_3^+$ flux contributes to 91% of the total ion flux, $H_2^+$ flux to 8% and $H^+$ to 11% in the ECR plasma, while in the RF plasma $H_3^+$ flux forms around 66% of the total ion flux, $H_2^+$ flux – 26% and $H^+$ – 8%.

An important remark has to be made about hyperthermal H atoms, which are created through the electron impact dissociation of $H_2$ in plasma with energies of few eV per H [31,32]. In the present RF-based NI sources for fusion reactors the grid covered with Cs has a potential close to the plasma potential, hence most of hydrogen ions and hyperthermal neutrals have more or less the same energy when they strike the cesiated Mo surface [33]. Given a much larger H flux compared to a positive ion flux in these sources [34], the dominant mechanism of $H^-$ formation is conversion of hyperthermal H atoms on the Cs surface [1,8,32]. When a hydrogen atom approaches the metal surface its affinity level is downshifted by the image potential, which eases the tunneling transfer of an electron from the surface to form $H^-$. However, as soon as $H^-$ leaves the surface, at a given distance its affinity level comes into resonance with empty states in the conduction band and the electron is lost back to the metal, unless the NI velocity is sufficiently high. This distance is larger for the low work-function metal, which reduces the probability for the affinity level electron to tunnel back to the material. Hence, slow $H^-$ ions produced from hyperthermal atoms survive only if the work function of the material is low enough, such as for the cesiated Mo surface (< 2 eV [32]). In case of materials studied here (HOPG or Gd), the "work function" is higher and $H^-$ produced from hyperthermal atoms do not contribute to the measured NIEDF. It has also been shown previously [4] that the flux of measured NI is proportional to the flux of positive ions bombarding the HOPG surface in our set-up.

## 3. Modelling principles

A model has been previously developed to obtain the NIEADF on the sample from the NIEDF measured by the mass spectrometer [24]. The main principle was to choose *a priori* the NIEADFs *f(E,θ)* and to validate *a posteriori* this choice. In this model the NI trajectories between the sample and the mass spectrometer are computed based on their initial energy $E$ and the angle of emission $\theta$. The sheaths in front of the sample and in front of the mass spectrometer can be considered planar in our experimental conditions [20]. The NI paths in the sheaths are computed in accordance with the potential variation given by the Child Langmuir law. The input parameters for the trajectory calculations, such as the electron density, the electron temperature, the plasma potential and the applied surface bias, are taken from the experiment (see Section 2, "Experimental set-up"). Those negative ions originating from the sample that miss the sampling orifice of the mass spectrometer or arrive to it with an angle $\theta_{MS}$ higher than the acceptance angle $\theta_{aa}$ are eliminated from the calculations. The energy distribution of the ions after they pass the MS entrance is labelled *f′(E)*. The acceptance angle of the mass spectrometer $\theta_{aa}$ is calculated using the SIMION software [35]; it slightly changes with the ion energy $E$ due to a chromatic aberration of the lens in the MS [29]. This variation is taken into account in the model. SIMION is also used to calculate the overall transmission of H$^-$ inside the MS: $T_{MS}(E)$. It drops down from 0.7 for an ion beam spread over the acceptance cone that arrives to the MS entrance at 130 eV to 0.1 at 230 eV. The energy distribution of NI at the MS detector *f″(E)* is calculated in the model by applying $T_{MS}(E)$ to *f′(E)*. Finally, *f″(E)* can be directly compared to the measured NIEDF *f*$_{exp}$*″(E)* in order to validate the choice of the initial *f(E,θ)* in the model.

The energy and angular distribution of backscattered and sputtered particles computed by the SRIM code has been chosen as the initial guess for *f(E,θ)*; a good agreement between the measured NIEDF *f*$_{exp}$*″(E)* and the modelled one *f″(E)* has been shown in case of HOPG [21,24]. The sample material is assumed to be an amorphous a-C:H layer (20% H), since the graphite surface exposed to plasma is subjected to hydrogen implantation and defect creation in the subsurface layer, which has been confirmed by Raman spectroscopy measurements [36,37]. The parameters of the SRIM calculation are listed elsewhere [21,24].

It is important to take into account all three populations of hydrogen ions present in plasma, $H_3^+$, $H_2^+$ and $H^+$, when calculating *f(E,θ)* with SRIM [24]. Here we extend the comparison of *f*$_{exp}$*″(E)* and *f″(E)* to the case of realistic PIEDFs (Figure 2), which were used as input to the SRIM code (in [24] the PIEDFs were not measured and monoenergetic ions were assumed). Figure 3 shows a comparison between the measured NIEDF (blue) in the ECR plasma and the computed one ($3·10^7$ incident ions in computation: red curve for *f′(E)* and green curve for *f″(E)*). In order to measure precisely the NIEDF over 5 orders of magnitude, the MS has been operated in the count accumulation regime for 20 min; it has been checked that the shape of NIEDF is not changing with time. In Figure 3 *f*$_{exp}$*″(E)* has been shifted by 1 eV to the left to match the peak of *f″(E)* given by the model.

The model reproduces quite well the shape of *f*$_{exp}$*″(E)*, which presents a main peak at low-energy (0–10 eV), a tail with a slight decreasing slope at intermediate energy (10–30 eV) and breakings of the slope around 30 eV and 50 eV followed by the high energy tail. Each change of the slope corresponds to a certain hydrogen ion population, as can be seen in Figure 3; this is also evident for the initial *f(E)* given by SRIM (black curve). Indeed, the energy of the ejected particle cannot exceed the maximum impact energy of the corresponding positive ion, see Figure 2; that is why there are 3 steps in *f(E)*. The experimental NIEDF is changing more smoothly in the range of ion energies 30–40 eV compared to the modelled one; this probably indicates a small inaccuracy in the determination of input PIEDF. It is possibly due to a variation of the MS transmission with ion energy and mass for $H^+$, $H_2^+$, $H_3^+$ or due to a contribution of fast H atoms created in the sheath, which is neglected here, see Section 2. We use the raw measured PIEDF as an input for modelling, so we do not aim for the perfect matching of the curves but rather to compare the tendencies: all the slope changes, the low-energy part and the high-energy tail are reproduced remarkably well by the model. Besides, let us note that the modelled $H_3^+$ contribution provides a good agreement of *f ″(E)* with $f_{exp}″(E)$ for 95% of the negative-ion population, as only few negative-ions have energy higher than 35 eV.

Figure 4 shows a comparison between $f_{exp}″(E)$ and the computed *f ″(E)* for $3·10^7$ incident ions in case of the RF plasma. A good agreement between the modelled *f ″(E)* and $f_{exp}″(E)$ is demonstrated, although the experimental NIEDF again appears to be smoother than the modelled one. The slope changing of *f(E)*, *f ′(E)* and *f ″(E)* is less pronounced in the RF plasma compared to the ECR case, as the PIEDF in the RF plasma has a smeared shape due to the ion collisions in the sheath.

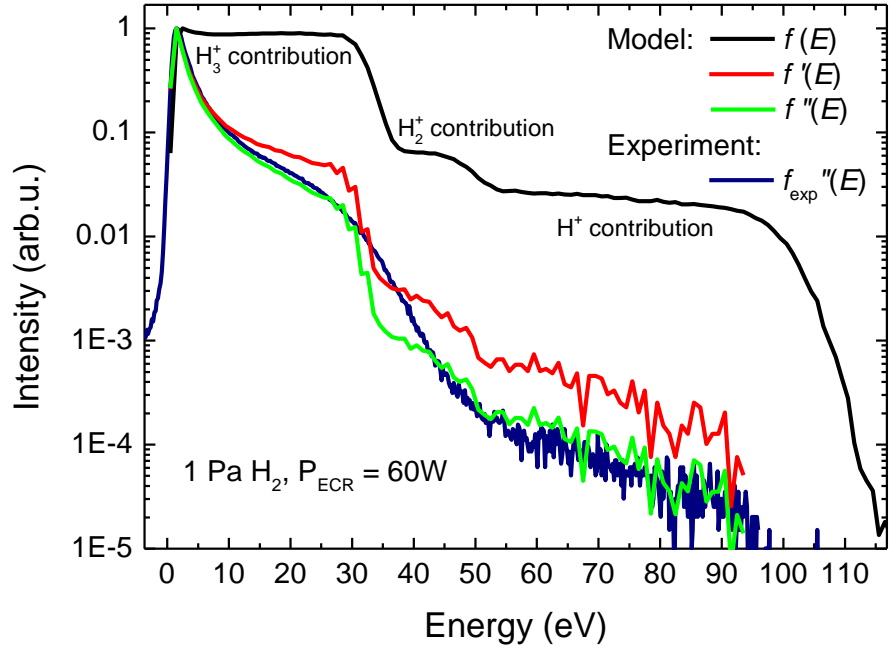

Figure 3. Comparison between the calculated energy distribution function *f ″(E)* of the negative ions at the MS detector (green line) and the experimental one $f_{exp}″(E)$ (blue line) obtained at 1 Pa, 60 W with the ECR source, $α = 0°$. The energy distribution function of ions on the sample *f(E)* calculated by SRIM and used as input in the model is shown with a black line. Red line shows *f ′(E)* at the MS entrance. All NIEDFs are normalized to the peak value.

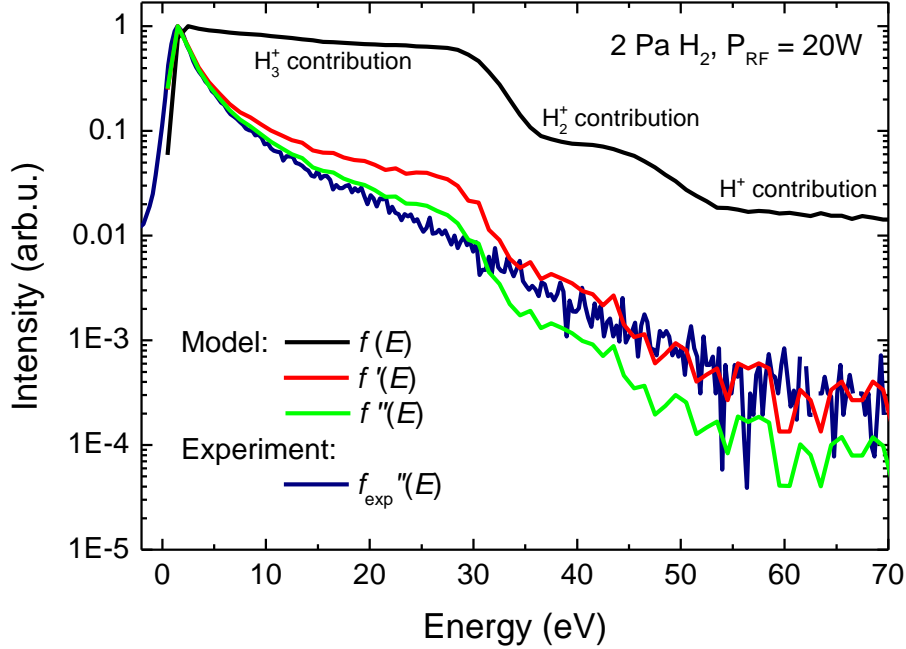

Figure 4. Comparison between the calculated energy distribution function $f''(E)$ of the negative ions at the MS detector (green line) and the experimental one $f_{exp}''(E)$ (blue line) obtained at 2 Pa, 20 W with the RF source, $\alpha = 0°$. The energy distribution function of ions on the sample $f(E)$ calculated by SRIM and used as input in the model is shown with a black line. Red line shows $f'(E)$ at the MS entrance. All NIEDFs are normalized to the peak value.

Finally, Figure 3 and Figure 4 demonstrate that only a part of the emitted ions is collected by the mass-spectrometer: the distribution function of the collected ions $f''(E)$ differs strongly from the distribution function of the emitted ions $f(E)$.

SRIM does not take into account the surface ionization, hence the previous modelling approach implicitly assumes that the surface ionization probability is independent of the angle and energy of the emitted particles. This might be true for carbon materials [21,24], but in case of metals the ionization probability is usually dependent on the outgoing velocity [26,38-40]. Moreover, the input parameters for SRIM may be unknown in general case. Therefore, it is crucial to develop a new modelling method to reconstruct the real NIEADF $f(E,\theta)$ of the negative ions leaving the sample surface based on the MS measurements.

## 4. Reconstruction method

In order to be collected, a negative-ion must arrive to the mass spectrometer sampling orifice with an angle $\theta_{MS}$ lower than the acceptance angle $\theta_{aa}$. This limits the collection of NI to those which have been emitted in a certain angular range $[\theta_1; \theta_2]$. This angular range depends on the energy of the emitted ions $E$ and also on the tilt angle between the sample normal and the MS axis $\alpha$ (see Figure 1). This was demonstrated in [24] and is shown in Figure 5: the segments of different colours correspond to the ranges of emission angles $[\theta_1; \theta_2]$ for which the ion is collected by the MS at a given $\alpha$ if emitted at energy $E$. One can see that by changing $\alpha$ at fixed $E$ the whole range of emission angles $\theta \in [0°; 90°]$ can be scanned. Therefore, it is possible to reconstruct the whole angular distribution of NI leaving the surface based on the NIEDF measurements at different tilt angles of the sample.

NIEDFs are measured for $\alpha = 0°$ to $35°$ with a step of $1°$. The step of $1°$ is a technical limitation of the experiment and above $35°$ the signal over noise ratio is too bad. One can see in Figure 5 that the ranges of $\theta$ which correspond to adjacent $\alpha$ overlap for a given energy. It means that a reconstruction method has to be developed in order to account for the correct contributions to the measured distribution $f_{exp}''(E)$ of NI having initial parameters $E$, $\theta$. In practice, mass-spectrometer collection efficiency needs to be calculated for each NI energy and angle. All negative-ions emitted by the sample are characterized by several parameters: their initial energy $E$, the direction of their velocity vector given by two angles in spherical coordinate system: $\theta$ (polar angle with respect to the sample normal) and an azimuthal angle $\varphi$, as well as their starting position on the sample surface plane given in a polar coordinate system by a radius $r$ and an azimuthal angle $\phi$ (see Figure 6). Let us note first that the NI emission is distributed uniformly on the surface: there is no dependence of emission on $r$ and $\phi$. Secondly, there is no preferred direction of emission in $\varphi$ and the emission distribution in $\theta$ angle is dependent on the emission mechanisms: backscattering, sputtering and ionization probability. The probability for a negative ion to leave the surface with an energy $E$ and a velocity vector at the polar angle $\theta$ is defined as $p(E,\theta)$.

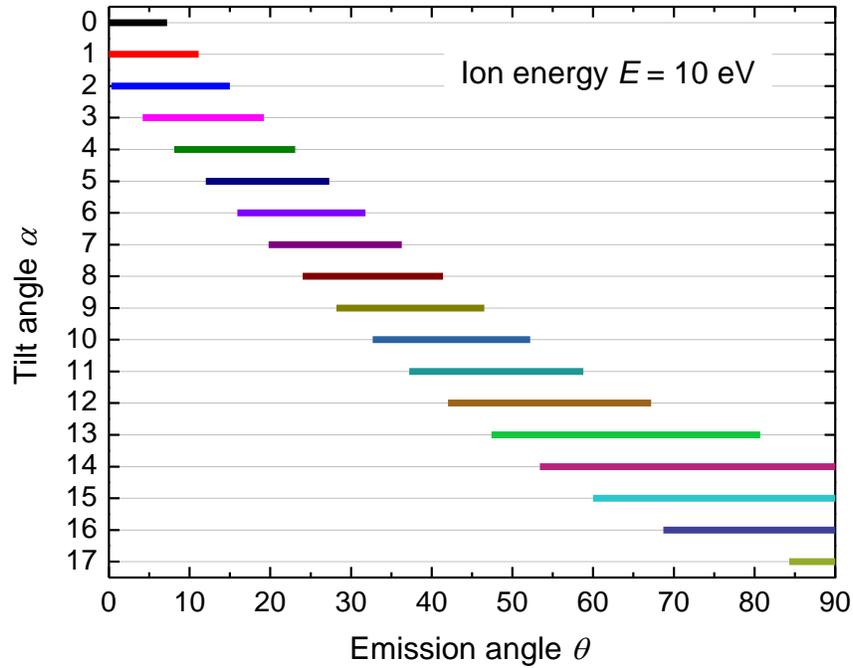

Figure 5. Ranges of polar emission angle $\theta$ which correspond to collected NI with initial energy $E = 10$ eV for different tilt angles $\alpha$ of the sample. The plot is calculated for the ECR $H_2$ plasma at 1 Pa, 60 W, $\theta_{aa} = 2°$.

If the NI flux emitted by the sample is defined as $\Gamma_0$ [ion/m$^2$/s], then the angular and energy distribution function of NI leaving the sample is $f(E,\theta) = \Gamma_0 \cdot p(E,\theta)$ and the fraction of flux emitted between $\theta$ and $\theta + d\theta$ and between $\varphi$ and $\varphi + d\varphi$ at energy $E$ is

$$dF = \frac{1}{2\pi} f(E,\theta) d\theta d\varphi \qquad (1)$$

The factor $2\pi$ is introduced for the normalization of the total distribution to $\Gamma_0$. All ions leaving the surface with a given set of $E$, $\theta$, $\varphi$ have paralell trajectories, see Figure 6b. Only those reaching the mass spectrometer have a chance to be collected. Those ions are coming from a small area on the sample surface which is the projection of the surface of the MS orifice $S_{MS}$ on the sample plane: $\Delta S = S_{MS}/\cos \alpha$ (Figure 6b), so that the fraction of ion flux (in units of ions/s) reaching the MS orifice for given emission angles $\theta$, $\varphi$ and energy $E$ is

$$dI_o = \frac{1}{2\pi} \frac{S_{MS}}{\cos \alpha} f(E,\theta) d\theta d\varphi. \qquad (2)$$

In order to obtain the measured intensity, i.e. the fraction of ion flux that reaches the MS detector, the transmission function of the MS $T_{MS}(E)$ must be taken into account:

$$dI = \frac{1}{2\pi} \frac{S_{MS}}{\cos \alpha} T_{MS}(E) f(E,\theta) d\theta d\varphi. \qquad (3)$$

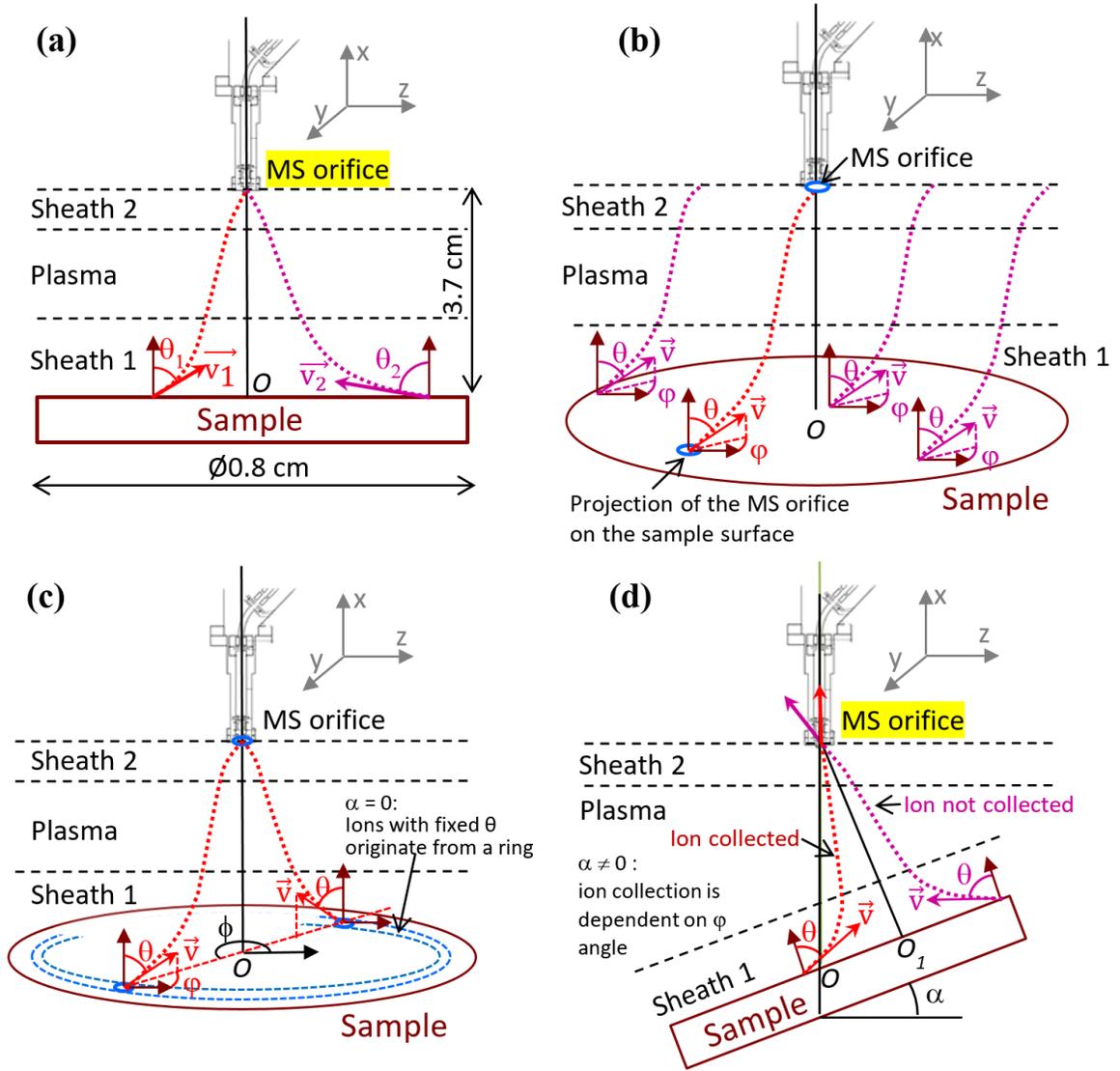

Figure 6. Scheme showing trajectories of negative ions emitted by a sample surface element $\Delta S = S_{MS} / \cos \alpha$. The drawing is not on scale: the sample size is enlarged, in reality most of NI are collected from a 2 mm diameter spot on the surface. The negative ion is collected if it enters the MS sampling orifice with a polar angle $\theta_{MS} < \theta_{aa}$ and if its deviation from the sample center $O$ is smaller than the radius of the sample. $\Delta S$ is located on a quasi-elliptic shape which is a geometrical locus of origins of ions emitted at a fixed polar angle $\theta$ and reaching the MS orifice. By varying $\theta$ one obtains a set of concentric quasi-elliptic shapes with a common center in $O_1$, which corresponds to $\theta = 0°$. $O_1$ does not coincide with $O$ if $\alpha \neq 0$.

At $\alpha = 0$, for a given $\varphi$ angle the element $\Delta S$ is located on the sample surface at a position $(r; \phi)$. The value of $r$ depends on $\theta$ and $E$ (Figure 6a) while the value of $\phi$ is directly given by $\varphi$: $\phi = \varphi + \pi$, as illustrated in Figure 6c. The ions originating from the element $\Delta S$ with a set of parameters $E$, $\theta$, $\varphi$ inevitably reach the MS orifice and be collected only if their arrival angle $\theta_{MS}$ is smaller than $\theta_{aa}$; if this condition is satisfied for one value of $\varphi$, ions emitted at $E$, $\theta$ with any $\varphi$ angle can be collected. Indeed, if $\varphi$ is varied at fixed $E$ and $\theta$ the element $\Delta S$ follows a circular ring on the sample surface, as shown in Figure 6c.

When $\alpha \neq 0$, the symmetry is broken, see Figure 6d. Firstly, the ring corresponding to origins of ions with fixed $E$, $\theta$ and $\varphi \in [0; 2\pi)$ that reach the MS entrance is displaced with respect to the MS axis; in fact, this ring is no longer circular, but rather quasi-elliptic. The ring would be perfectly circular if there was no deviation of ions trajectories in Sheath 2. Secondly and most importantly, the NI leaving the surface of such ring with different $\varphi$ do not arrive with the same angle to the MS sampling orifice, which implies that only a part of them can be collected, see Figure 6d. Therefore, at $\alpha \neq 0$, for each set of emission parameters $E$, $\theta$ only ions with $\varphi$ within the acceptable range [$\varphi_{min}$, $\varphi_{max}$] are collected. In consequence, the total ion flux measured by the MS for a given ion energy $E_m$ at a given tilt angle $\alpha_l$ can be written as follows:

$$I(E_m, \alpha_l) = \frac{1}{2\pi} \frac{S_{MS}}{\cos \alpha_l} T_{MS}(E_m) \int_{\theta_{min}(E_m,\alpha_l)}^{\theta_{max}(E_m,\alpha_l)} \int_{\varphi_{min}(E_m,\alpha_l,\theta)}^{\varphi_{max}(E_m,\alpha_l,\theta)} f(E_m, \theta) d\theta d\varphi. \quad (4)$$

Here [$\theta_{min}(E_m,\alpha_l)$; $\theta_{max}(E_m,\alpha_l)$] and [$\varphi_{min}(E_m,\alpha_l,\theta)$, $\varphi_{max}(E_m,\alpha_l,\theta)$] are the ranges of emission angles for which the ions are collected by MS at a given $\alpha_l$.

Let us introduce the collection efficiency matrix $K(E_m,\alpha_l,\theta)$ in the following way:

$$K(E_m, \alpha_l, \theta)$$
$$= \begin{cases} \dfrac{1}{2\pi} \dfrac{1}{\cos \alpha} \displaystyle\int_{\varphi_{min}(E_m,\alpha_l,\theta)}^{\varphi_{max}(E_m,\alpha_l,\theta)} d\varphi, & \text{if } \theta \in [\theta_{min}(E_m, \alpha_l); \theta_{max}(E_m, \alpha_l)] \\ 0, & \text{if } \theta \notin [\theta_{min}(E_m, \alpha_l); \theta_{max}(E_m, \alpha_l)] \end{cases}$$
$$= \begin{cases} \dfrac{1}{\cos \alpha_l} \dfrac{\varphi_{max}(E_m, \alpha_l, \theta) - \varphi_{min}(E_m, \alpha_l, \theta)}{2\pi}, & \text{if } \theta \in [\theta_{min}(E_m, \alpha_l); \theta_{max}(E_m, \alpha_l)] \\ 0, & \text{if } \theta \notin [\theta_{min}(E_m, \alpha_l); \theta_{max}(E_m, \alpha_l)] \end{cases} \quad (5)$$

Then $I(E_m,\alpha)$ becomes

$$I(E_m, \alpha_l) = S_{MS} T_{MS}(E_m) \int_0^{90°} K(E_m, \alpha_l, \theta) f(E_m, \theta) d\theta. \quad (6)$$

In other words the measured intensity $I(E_m,\alpha_l)$ corresponds to the integral of unknown angular distribution function $f(E_m,\theta)$ weighted with the collection efficiency matrix $K(E_m,\alpha_l,\theta)$. The latter is calculated in the following way: a uniform ion distribution in $E$, $\theta$ and $\varphi$ is introduced for each $\alpha$. Then the model calculates the ion trajectories and checks for each combination $E$, $\theta$, $\alpha$ and $\varphi$ if the ion is collected in order to determine $\varphi_{max}(E_m, \alpha_l, \theta)$ and $\varphi_{min}(E_m, \alpha_l, \theta)$. From a practical point of view the collection efficiency is given by discretizing Eq. (5):

$$K(E_m, \alpha_l, \theta_j) = \frac{1}{\cos \alpha_l} \frac{\sum_{i=1}^{P} n_{\text{collect}}(E_m, \alpha_l, \theta_j, \varphi_i)}{P} \quad (7)$$

Here $n_{\text{collect}}(E_m, \alpha_l, \theta_j, \varphi_i) = 1$ if ion is collected and 0 otherwise, $\theta$ goes from 0° to 90° with a step $\Delta\theta = 0.3°$, $\varphi$ goes from 0° to 359° with a step $\Delta\varphi = 1°$ (normalization factor $P = 360$). Energy $E_m$ goes from 1 eV to 50 eV with the discretization step of 0.2 eV. Example of $K(E_m, \alpha, \theta)$ for the initial ion energy $E_m = 10$ eV is shown in Figure 7. Lines of different colors correspond to different $\alpha$.

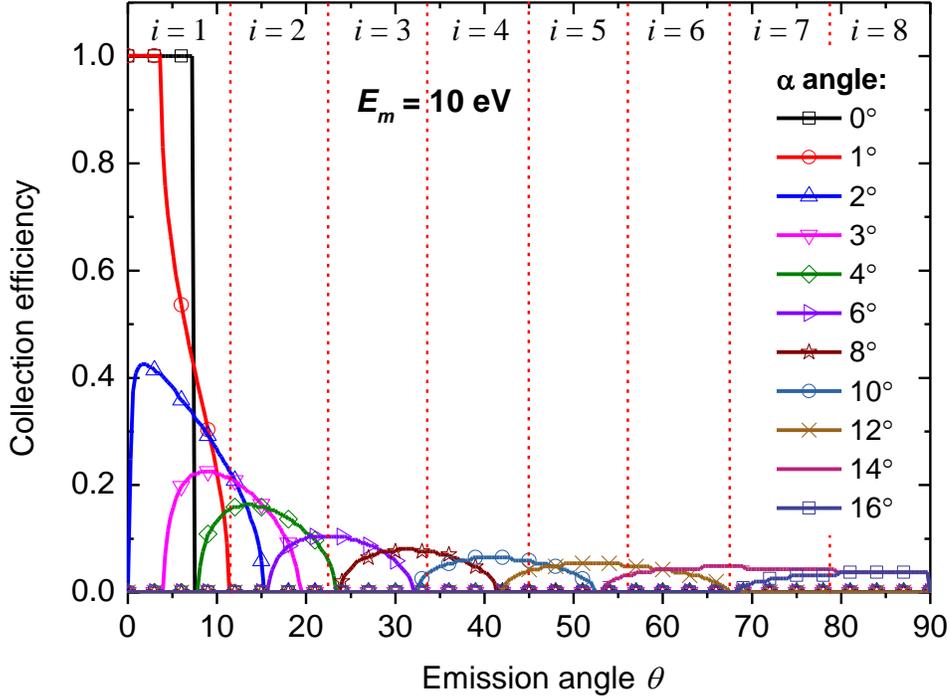

Figure 7. Collection efficiency matrix $K(E_m, \alpha, \theta)$ calculated for the ion energy $E_m = 10$ eV in case of the ECR $H_2$ plasma at 1 Pa, 60 W, $\theta_{aa} = 2°$.

The complete range of $\theta$ from 0° to 90° is divided into $N = 8$ uniform intervals ($N = 8$ has been chosen for the reasons described below). The sought function $f(E_m, \theta)$ is assumed to be a second order polynomial function of $\theta$ on each of these intervals, with continuity conditions between intervals that are described hereafter. The intervals are marked by red dashed lines in Figure 7. Now Eq. (6) can be discretized in the following way

$$I(E_m, \alpha_l) = S_{MS} T_{MS}(E_m) \sum_{i=1}^{N} \sum_{j=j_{\min}(i)}^{j_{\max}(i)} K(E_m, \alpha_l, \theta_j) f_i(E_m, \theta_j) \Delta\theta \qquad (8)$$

with a set of boundary conditions:

$$\begin{cases} f_i(E_m, \theta_{\max}^i) = f_{i+1}(E_m, \theta_{\max}^i) & (a) \\ f_i'(E_m, \theta_{\max}^i) = f_{i+1}'(E_m, \theta_{\max}^i) & (b) \\ f_N(E_m, \theta_{\max}^N = 90°) = 0 & (c) \\ f_N'(E_m, \theta_{\max}^N = 90°) = \delta & (d) \end{cases} \qquad (9)$$

Here $l = 1…L$ is the index of $\alpha$, $i = 1…N$ is the index of the interval (see Figure 7, $f_i$ is defined by a second order polynomial function on each interval $i$), $j$ is the index of $\theta$ which goes from $j_{min}(i)$ to $j_{max}(i)$ that correspond to the boundaries of the interval $i$: $\theta \in \left[\theta_{j_{min}(i)}; \theta_{j_{max}(i)}\right]$ or $\theta \in [\theta^i_{min}; \theta^i_{max}]$, $f'(E_m, \theta) = df/d\theta$ and we assume continuity of both the function $f$ and its derivative between intervals (Eq. 9a and b). It is assumed that no ions are emitted parallel to the sample surface, i.e. $f(E, \theta = 90°) = 0$ (Eq. 9c). Finally, the slope of the distribution function at $\theta = 90°$, $f'(\theta = 90°)$ is a free parameter $\delta$ of the system (Eq. 9d).

In order to linearize the system of equations (8) we assume that the sought angular distribution function is parabolic on each interval $i$: $f_i(E_m, \theta_j) = k_i\theta_j^2 + b_i\theta_j + c_i$. Such hypothesis allows to find a smooth solution $f(E_m, \theta)$ on the whole range of $\theta$ by matching $f_i(E_m, \theta_j)$ and $f_i'(E_m, \theta_j)$ correspondingly on the boundaries of the adjacent intervals, see Eq. (9). Hence Eqs. (8) and (9) are rewritten in the form of a system of linear equations:

$$\begin{cases} \dfrac{I(E_m, \alpha_l)}{S_{MS}T_{MS}(E_m)} = \sum_{i=1}^{N} \sum_{j=j_{min}(i)}^{j_{max}(i)} K(E_m, \alpha_l, \theta_j)[k_i\theta_j^2 + b_i\theta_j + c_i]\Delta\theta \\ k_i(\theta^i_{max})^2 + b_i\theta^i_{max} + c_i = k_{i+1}(\theta^i_{max})^2 + b_{i+1}\theta^i_{max} + c_{i+1} \\ 2k_i\theta^i_{max} + b_i = 2k_{i+1}\theta^i_{max} + b_{i+1} \\ k_N(\theta^N_{max})^2 + b_N\theta^N_{max} + c_N = 0 \\ 2k_N\theta^N_{max} + b_N = \delta \end{cases} \quad (10)$$

The whole system (10) is written in the matrix form and solved to find unknowns $k_i$, $b_i$ and $c_i$. The matrix dimensions are $2N + L$ by $3N$. The number $L$ of tilt angles $\alpha$ is determined by experimental considerations (signal to noise ratio). It is clear that the number of intervals $N$ must be less or equal to $L$, otherwise the system is underdetermined. In practice, $N = \min(L - 2, 8)$ is chosen to obtain an overdetermined system, which improves stability of the solution. The problem is solved with the MATLAB "*lsqlin*" function, which solves the linear system $Fx = d$ in the least-squares sense (by minimizing the squared norm $\|Fx - d\|_2$) with linear inequality constraints. The latter is useful to avoid negative unphysical solutions. By changing the free parameter $\delta$ (see Eq. (10)) the best solution is found, i.e. without numerical oscillations and with the lowest norm $\|Fx - d\|_2$. The solution is the angular distribution function at a given energy $E_m$: $f(E_m, \theta)$. It is sensitive to the noise in the experimental values $I(E_m, \alpha_l)$. Hence the following smoothing procedure is applied: first the NIEDF is averaged over three adjacent energies centered on $E_m$ in order to get $I(E_m, \alpha_l)$. The procedure is repeated for all $\alpha_l$. Then a linear interpolation of $I(E_m, \alpha)$ with a step $\Delta\alpha = 0.1°$ and smoothing with $2^{nd}$ order Savitzky-Golay filter with Gaussian window of 35 points is applied. Such approach eliminates noisy oscillations from the measured intensity while keeping the general tendency. The algorithm described above is applied for all ion energies $E$ from 1 to 50 eV with a step of 0.2 eV. Hence the complete NIEADF $f(E, \theta)$ is reconstructed from the experimental data. Such method does not use any *a priori* assumption on $f$ or on the ionization probability; therefore it can be applied to any type of material and NI.

## 5. Application to HOPG

The reconstruction method is firstly applied to the HOPG sample exposed to the $H_2$ plasma; the ECR regime is chosen due to higher NI intensities and less perturbed NIEDF measurements compared to the RF plasma. The energy and angular distribution of neutrals leaving the HOPG surface bombarded with positive hydrogen ions (with PIEDF shown in Figure 2) predicted by the SRIM code is given in Figure 8a in the form of polar contour plot. The radial axis represents energy $E$, the angular one – the angle of emission $\theta$, the colour stands for the value of $f_{SRIM}(E,\theta)$. It can be seen that most of neutrals are ejected from the surface with energies below 30 eV and angles symmetrically distributed around $\theta = 45°$. It is possible to benchmark the reconstruction method presented above. First a direct model [24] is applied to the SRIM distribution producing NIEDFs seen by the MS at different tilts $\alpha$ of the sample (such as red curves in Figure 3 and Figure 4). Then the output of the direct model is introduced as an input to the reconstruction method, so that the reconstructed $f_{rec}(E,\theta)$ is calculated and compared to the initial $f_{SRIM}(E,\theta)$.

The result of the reconstruction $f_{rec}(E,\theta)$ is shown in Figure 8b demonstrating a good agreement with the original distribution $f_{SRIM}(E,\theta)$. The same may be concluded about the integrated distributions $f_{SRIM}(E)$ and $f_{rec}(E)$ in Figure 9a and $f_{SRIM}(\theta)$ and $f_{rec}(\theta)$ in Figure 9b; the original data given by SRIM is shown with black solid lines, while the reconstructed result is plotted with red dash-dot lines. These results demonstrate the consistency of both modelling methods. The noise introduced by the method in Figure 8b is due to the fact that there is no correlation between the solutions at neighbouring values of ion energy (except for averaging the input intensity $I(E_m,\alpha)$ over $E_m \pm 0.2$ eV).

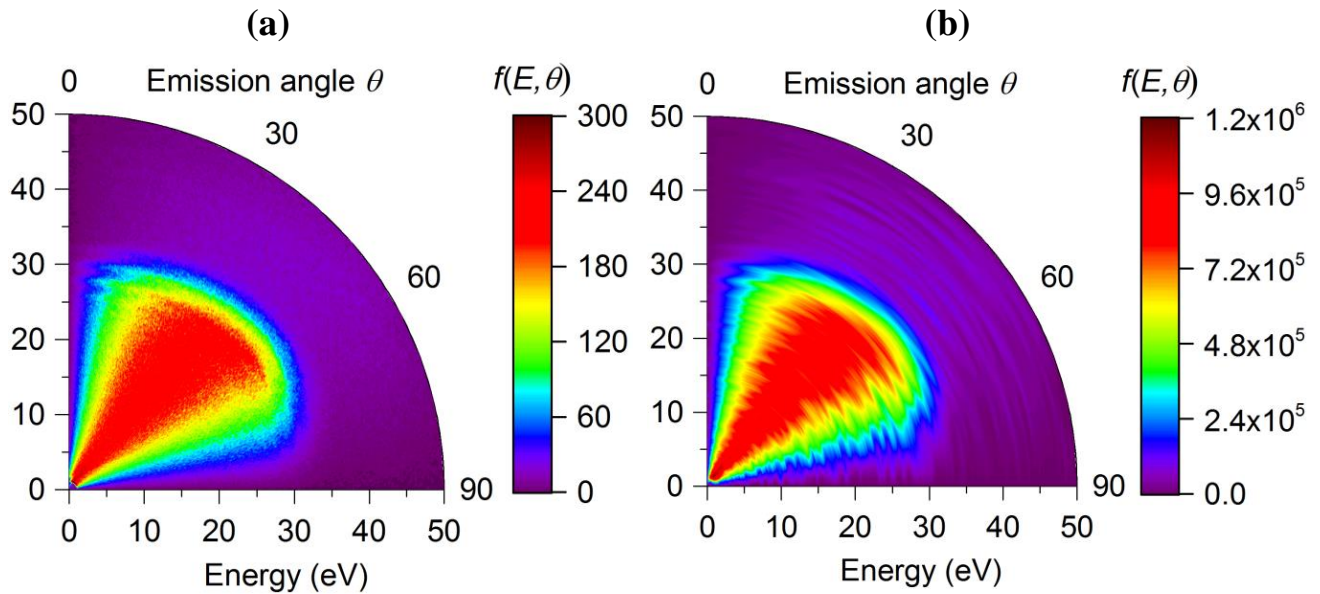

Figure 8. $f_{SRIM}(E,\theta)$ of the emitted particles given by SRIM for the HOPG surface (a) and its reconstruction $f_{rec}(E,\theta)$ from the direct model applied to the SRIM output (b).

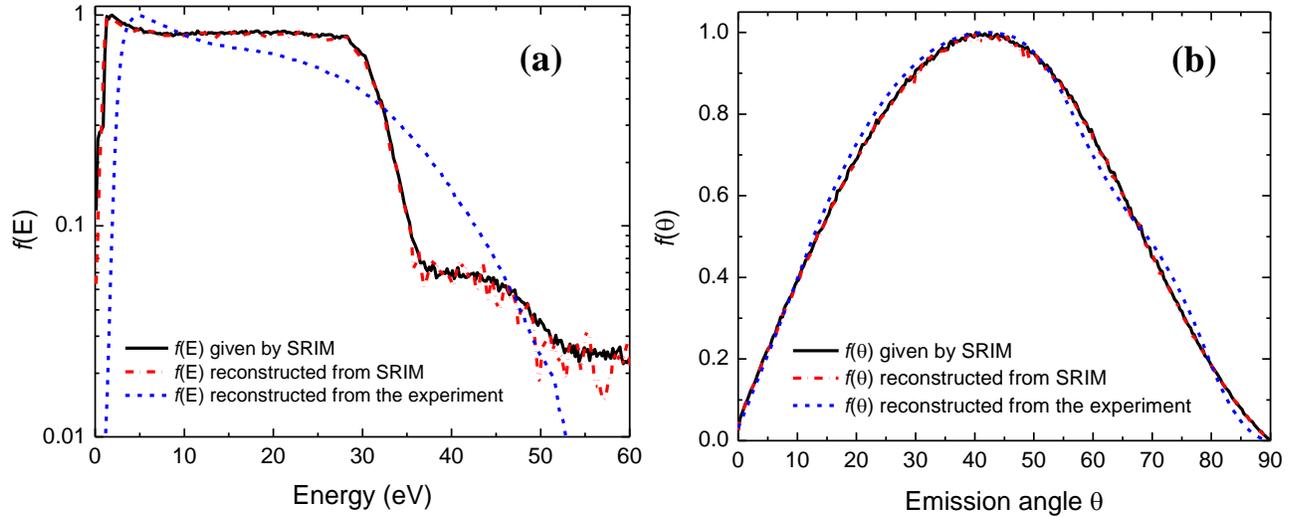

Figure 9. Normalized integrated distributions of the emitted particles: $f_{SRIM}(E)$ given by SRIM, reconstructed $f_{rec}(E)$ either from SRIM or from experiment (a) and $f_{SRIM}(\theta)$ given by SRIM, reconstructed $f_{rec}(\theta)$ either from SRIM or from experiment for the HOPG surface (b).

As far as the experiment is considered, Figure 10 shows measured NIEDFs for different tilts of the HOPG sample polarized at -130V. When $\alpha$ increases, the energy onset of the distribution shifts to higher energies. It can be seen that the curves perfectly match each other for the energies higher than the peak value. The resulting NIEADF $f_{rec}(E,\theta)$ given by the reconstruction method based on this data is shown in Figure 11.

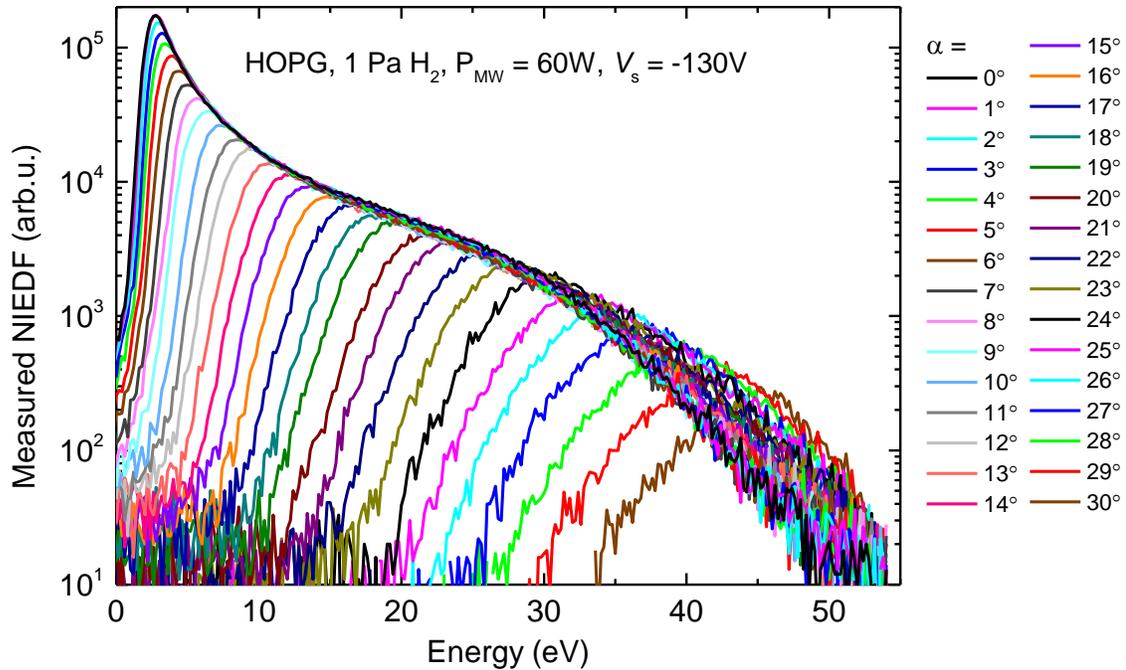

Figure 10. Experimentally measured NIEDF for different tilts of HOPG sample ($\alpha = 0° - 35°$ with a step of 1°) in case of the ECR $H_2$ plasma at 1 Pa, 60 W.

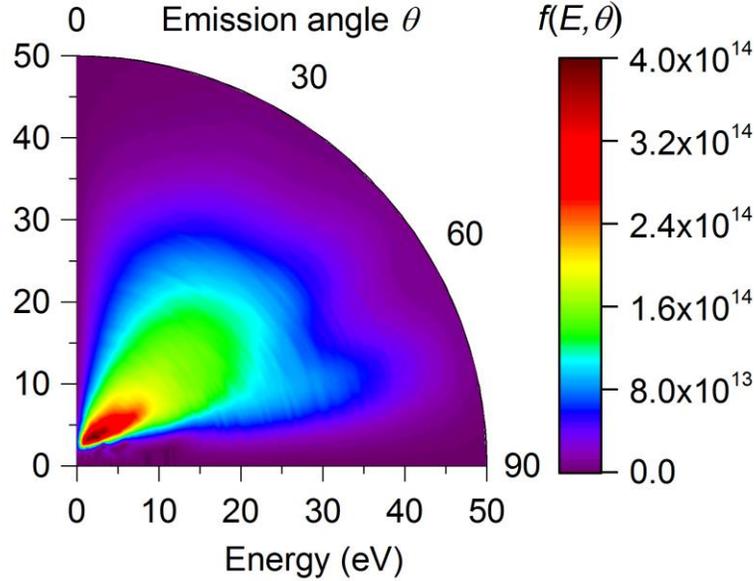

Figure 11. NIEADF obtained by the reconstruction method applied to the experimental data in case of the HOPG sample in the ECR $H_2$ plasma at 1 Pa, 60 W.

Although $f_{rec}(E,\theta)$ does not exactly match $f_{SRIM}(E,\theta)$ in Figure 8, their shapes are comparable: maximum outgoing energy of 30 eV and an overall symmetry around $\theta = 45°$. It is reminded that SRIM calculation does not take into account ionization of neutrals leaving the surface, while the reconstruction of experimental data is based on the direct measurements of negative ions collected by MS. The main difference between $f_{rec}(E,\theta)$ given by the reconstruction method and $f_{SRIM}(E,\theta)$ is that the majority of negative ions are created with energies less than 10 eV instead of 30 eV. This is evident in Figure 9a, where the integrated distributions $f_{rec}(E)$ and $f_{SRIM}(E)$ are shown: the SRIM code predicts a stepwise energy distribution, where each step corresponds to a different positive hydrogen ion in plasma as shown in Figure 2, while $f_{rec}(E)$ is much smoother. In addition, the low energy peak of $f_{rec}(E)$ is slightly shifted to the right: it corresponds to the peak on the experimental NIEDF at $\alpha = 0°$ in Figure 10. As far as the integrated angular distribution is considered (Figure 9b), $f_{rec}(\theta)$ is rather close to $f_{SRIM}(\theta)$ and their maxima coincide. The shapes of $f_{rec}(E,\theta)$ and $f_{SRIM}(E,\theta)$ agree qualitatively and $f_{SRIM}(\theta)$ is very close to $f_{rec}(\theta)$; the origin of discrepancies between $f_{SRIM}(E)$ and $f_{rec}(E)$ is probably the use of raw measured PIEDF as an input for SRIM modelling. As in Section 3, we do not aim for the perfect matching with SRIM but rather to compare the tendencies.

Considering that reconstructed $f_{rec}(E,\theta)$ represents a real distribution of the NI emitted from the surface, it has been used as an input for the direct model [24] for $\alpha = 0°$. The result is shown in Figure 12 ($f_{rec}(E)$ and $f_{rec}''(E)$ by black and red points correspondingly) and compared to the experimental data for HOPG in ECR plasma (blue curve). The agreement is perfect, which validates again the coherence of the direct NIEDF model and the reconstruction method.

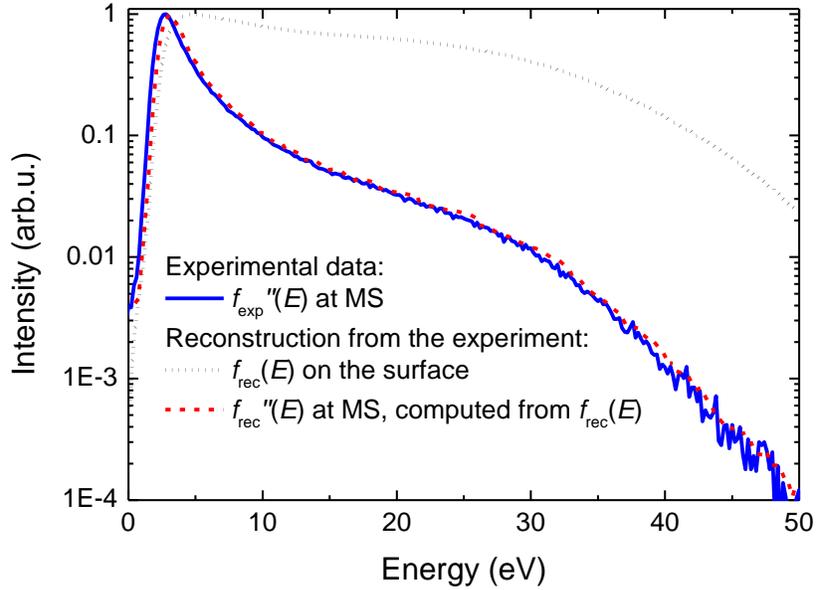

Figure 12. Comparison between experimental NIEDF (blue) and calculated NIEDF for HOPG in hydrogen. $f_{rec}(E)$ emitted from the surface given by reconstruction method is shown by black dots, $f_{rec}''(E)$ at the MS detector given by the direct model (the reconstructed distribution $f_{rec}(E,\theta)$ was used as input) is shown by red dots.

## 6. Application to Gadolinium

Gadolinium (Gd) sample with the purity of 99.9 % has been chosen as an example of a low work-function metal. It is reminded that the ionization probability $P_{iz}$ is probably dependent on the perpendicular velocity of the emitted ions [26,38-40]. The angle-resolved measurements of NIEDF performed on Gd are shown in Figure 13. Comparing to HOPG (Figure 10), one can see that the tail of the distribution is significantly higher than in case of HOPG. This is due to an increased contribution of backscattered particles with high energies. As Gd atoms are much heavier than carbon ones, the hydrogen momentum transfer to the bulk of the Gd material is less efficient and light hydrogen ions are more efficiently scattered by the gadolinium surface. The onset of NIEDF shifts with the tilt angle $\alpha$ and the tails superpose in the same way as for HOPG.

It is reminded that the SRIM simulation parameters for hydrogen interaction with Gd surface are not well established. We adopted parameters given by the software: 25 eV for displacement energy, 3.57 eV for surface binding energy and 3 eV for lattice binding energy. It is also assumed that the hydrogen surface coverage is negligible; indeed, sputtered particles have a lower average energy than the backscattered ones and would contribute only to the low energy part of the NIEDF, while the low-energy peak in Figure 13 is not pronounced. It should be noted that SRIM gave a strange unphysical result: $f_{SRIM}(E)$ was strongly oscillating, hence a smoothing with a 3 eV window had to be applied. The resulting NIEADF $f_{SRIM}(E,\theta)$ calculated by the SRIM code is given in Figure 14 in the form of a polar contour plot.

Given all these facts, we do not aim to compare directly SRIM output and reconstruction method, but rather to look at the tendencies. The NIEADF $f_{rec}(E,\theta)$ produced by the reconstruction method based on the experimental data (Figure 13) is shown in Figure 15. The agreement between the SRIM calculation and the distribution reconstructed from the experiment is unexpectedly good. There is a difference in the low-energy region ($E < 10$ eV) where $f_{rec}(E,\theta)$ reveals less NI than predicted by the SRIM model, which might be an artifact from the smoothing of $f_{SRIM}(E)$.

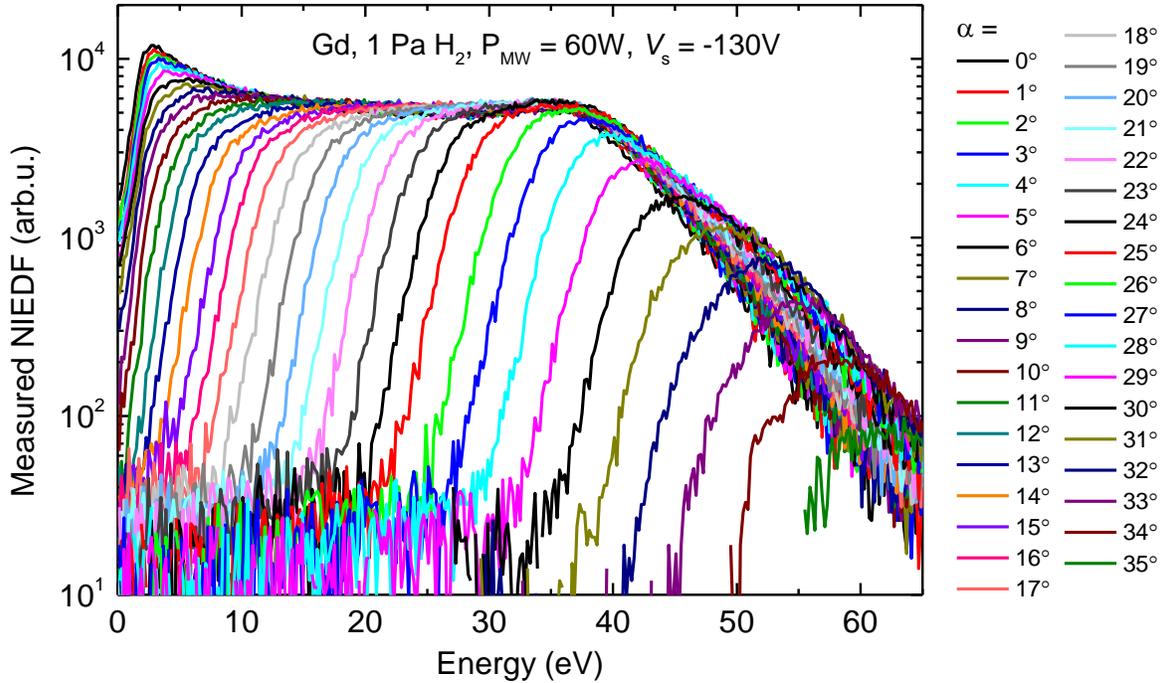

Figure 13. Experimentally measured NIEDF for different tilts of Gd sample ($\alpha = 0° - 35°$ with a step of 1°) in case of the ECR H$_2$ plasma at 1 Pa, 60 W.

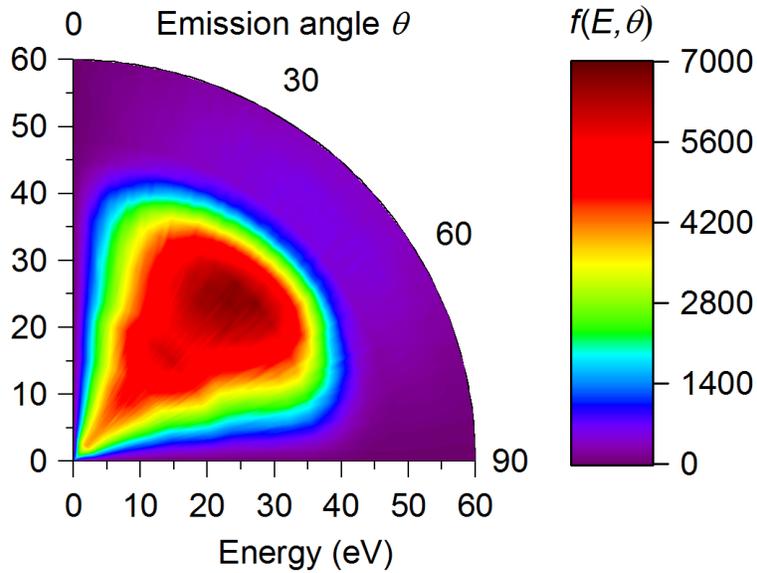

Figure 14. $f_{SRIM}(E,\theta)$ of the emitted particles given by SRIM for the Gd surface.

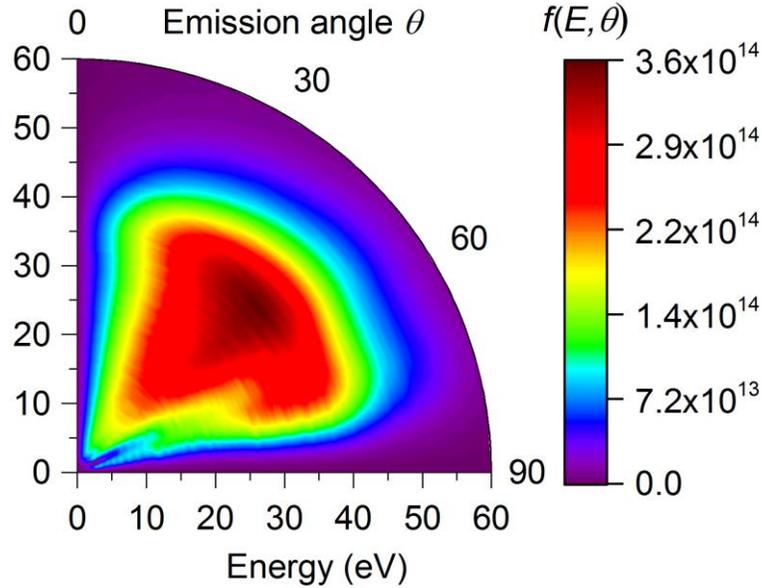

Figure 15. NIEADF obtained by the reconstruction method applied to the experimental data in case of the Gd sample in the ECR $H_2$ plasma at 1 Pa, 60 W.

The normalized integrated energy and angular distributions $f(E)$ and $f(\theta)$ are shown in Figure 16, both reconstructed and SRIM output. One can see that the $f_{rec}(E)$ distribution for Gd is considerably different from the one obtained for carbon (Figure 11). The energy distribution $f(E)$ for Gd reveals a substantial increase with energy up to the peak at 36 eV with a decrease at higher energies, while $f(E)$ for HOPG (Figure 9a) shows a low energy peak at 5 eV and a substantial decrease at higher energies. The tendency of increasing $f(E)$ up to 36 eV is predicted both by the experimental reconstruction and SRIM; even that the latter should not be directly applied if $P_{iz}(E,\theta)$ is unknown, such agreement reinforces our confidence in the method. In fact, the influence of the backscattering mechanism (which is included in SRIM) on the variation of negative-ion flux with negative-ion energy seems to dominate over the ionization mechanism (which is not taken into account in SRIM).

The effect of atomic hydrogen is more prominent for the HOPG surface (Figure 9a), where substantial H adsorption in the sub-surface layer can occur. In order to clarify the impact of the adsorbed hydrogen for Gd, we performed SRIM calculations with 10% and 20% of adsorbed H on Gd surface (surface binding energy of H is assumed to be 3 eV). As it can be seen in Figure 16a the presence of H on the surface does not change much the energy distribution of the outgoing particles. Moreover, adding H on the surface tends to move the low-energy part of $f_{SRIM}(E)$ further away from $f(E)$ reconstructed from the experiment.

The shoulder at 70° in the experimental $f(\theta)$ in Figure 16b could be artificial and draws attention to the present limits of the method: it originates from the fact that the range of $\alpha$ for which the measured intensities are non-zero at a given ion energy is larger than expected from the model. Since the ion trajectory calculation is very sensitive to the geometry, it means that the sample is not perfectly aligned with respect to the orifice of the MS in the experiment.

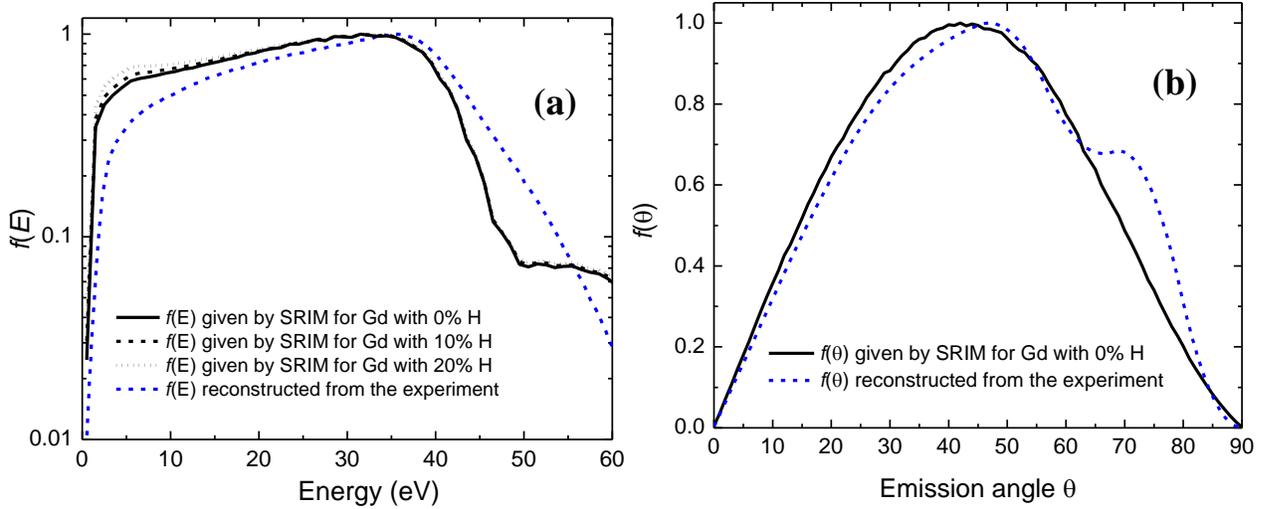

Figure 16. Normalized integrated distributions of the emitted particles: $f_{SRIM}(E)$ given by SRIM, reconstructed $f_{rec}(E)$ from experiment (a) and $f_{SRIM}(\theta)$ given by SRIM, reconstructed $f_{rec}(\theta)$ from experiment for Gd (b).

Finally, it is important to compare HOPG and Gd surfaces in terms of relative NI production efficiency. In view of this objective two experiments with exactly the same conditions and the same MS settings have been made both for Gd and HOPG: 1 Pa $H_2$, 60 W ECR power. The resulting reconstructed energy distributions $f_{rec}(E)$ with the same arbitrary units (based on the counts per second measured by the MS) are shown in Figure 17. The integrated NI yield produced by HOPG surface is 1.5 times higher than that of Gd; HOPG dominates in emitting NI with energies below 25 eV, while Gd is more efficient at higher energies of emitted ions. Although we cannot perform absolute yield calibrations in our experimental set-up, it is justified that HOPG surface is an efficient NI enhancer: despite being a semi-metal with high "work function" (4.5 eV [41]) it produces as much as the surface of a low work function metal (2.9 eV [27]). Therefore the reconstruction method allows to perform a relative comparison of the NI surface production efficiencies of different materials regardless of any assumptions about the mechanisms behind.

## 7. Conclusion

The experimental method combined with a proper model described in this paper provides a unique way to analyze the measured NIEDF and to obtain the initial NIEADF $f(E,\theta)$ of negative ions leaving the surface for any material. In fact, the information about the angles of emission, which is lost in the mass spectrometer, is recovered by tilting the sample.

A good overall agreement of the SRIM distribution $f_{SRIM}(E,\theta)$ with the one reconstructed from the experimental data $f_{rec}(E,\theta)$ verified our choice to use SRIM for the initial distribution $f_{SRIM}(E,\theta)$ for carbon materials, since the input parameters for SRIM calculation on a-C:H layers are well known in the scientific community. The reconstruction method is validated by a good agreement of SRIM calculations on carbon with the distributions reconstructed from experimental data for HOPG on several levels.

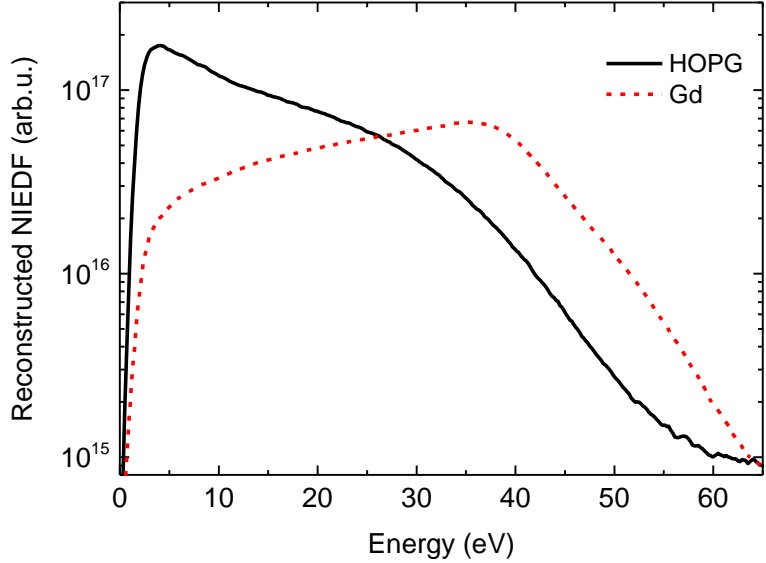

Figure 17. NIEDF on the sample surface $f_{rec}(E)$, reconstructed from the experiment performed in ECR $H_2$ plasma for HOPG (black solid) and Gd (red dashed) under the same conditions (1 Pa $H_2$, 60 W).

From a more general point of view, the obtained NIEADFs on two different materials such as HOPG and Gd prove that the reconstruction method can be used successfully for any material, even in case when the ionization probability $P_{iz}$ is not *a priori* constant with $E$ and $\theta$ of the emitted particle. The only input which is necessary to calculate the ion trajectories are the parameters of the plasma and the sheaths. In comparison with the previous method [24], the reconstruction method does not depend on the parameters of SRIM calculations, so it can be successfully applied to any type of surface and/or NI, especially when an *a priori* guess of the distribution function $f(E,\theta)$ is not available or the NI formation mechanism is not known.

Another important point is that HOPG shows a result comparable with the low work function metal Gd in terms of NI surface production, which justifies the choice of carbon materials as potential NI enhancers. The reconstruction method helps to identify the key differences in mechanisms of NI production: while for Gd the major process is backscattering of ions, in case of HOPG the sputtering contribution due to adsorbed H on the surface is also important. The reconstruction method can be applied to compare the relative NI surface production efficiencies of different materials, for instance, cesiated and non-cesiated surfaces. This is important for a development of new NI sources for fusion based on Cs-free materials.

## Acknowledgments

This work was carried out within the framework of the French Research Federation for Fusion Studies (FR-FCM) and the EUROfusion Consortium and received funding from the Euratom research and training programme 2014-2018 under grant agreement No 633053. The views and opinions expressed herein do not necessarily reflect those of the European Commission. Financial support was also received from the French Research Agency (ANR) under grant 13-BS09-0017 H INDEX TRIPLED.